\documentclass[12pt]{iopart}
\usepackage{amssymb,amstext}
\usepackage{graphicx}
\usepackage{bm}
\usepackage[colorlinks]{hyperref}
\begin{document}
\title{Supernova neutrinos and nucleosynthesis}
\author{G. Mart\'inez-Pinedo$^{1,2}$, T. Fischer$^{3}$, L. Huther$^{1}$} 
\address{$^1$Institut f{\"u}r
  Kernphysik (Theoriezentrum), Technische Universit{\"a}t Darmstadt,
  Schlossgartenstra{\ss}e 2, 64289 Darmstadt, Germany}
\address{$^2$GSI Helmholtzzentrum f\"ur Schwerioneneforschung,
  Planckstra{\ss}e~1, 64291 Darmstadt, Germany}
\address{$^3$Institute for Theoretical Physics, University of Wroc{\l}aw,
  pl. M. Borna 9, 50-204 Wroc{\l}aw, Poland} 
\ead{gabriel.martinez@physik.tu-darmstadt.de}

\begin{abstract}
  Observations of metal-poor stars indicate that at least two
  different nucleosynthesis sites contribute to the production of
  r-process elements. One site is responsible for the production of
  light r-process elements $Z\lesssim 50$ while the other produces
  the heavy r-process elements. We have analyzed recent observations
  of metal-poor stars selecting only stars that are enriched in light
  r-process elements and poor in heavy r-process elements. We find a
  strong correlation between the observed abundances of the $N=50$
  elements (Sr, Y and Zr) and Fe. It suggest that neutrino-driven winds from
  core-collapse supernova are the main site for the production of
  these elements. We explore this possibility by performing
  nucleosynthesis calculations based on long term Boltzmann neutrino
  transport simulations. They are based on an Equation of State that
  reproduces recent constrains on the nuclear symmetry energy. We
  predict that the early ejecta is neutron-rich with $Y_e \sim 0.48$,
  it becomes proton rich around 4~s and reaches $Y_e = 0.586$ at 9~s
  when our simulation stops. The nucleosynthesis in this model
  produces elements between Zn and Mo, including $^{92}$Mo. The
  elemental abundances are consistent with the observations of the
  metal-poor star HD~12263.  For the elements between Ge and Mo, we
  produce mainly the neutron-deficient isotopes. This prediction can
  be confirmed by observations of isotopic abundances in metal-poor
  stars. No elements heavier than Mo ($Z=42$) and no heavy r-process
  elements are produced in our calculations. 
\end{abstract}
\submitto{\jpg}
\maketitle

\section{Introduction}
\label{sec:introduction}

Core-collapse supernova occur at the end of the evolution of massive
stars when the core collapses to form a proto neutron star
(PNS)~\cite{Bethe:1990,Janka.Langanke.ea:2007}. The energy gain during
the collapse corresponds to the gravitational binding energy of the
PNS, $\approx 3\times 10^{53}$~ergs, and is emitted as neutrino
radiation on time scales of tens of seconds during which the central
PNS cools, deleptonizes and contracts to the final neutron star. This
corresponds to an emission of $10^{58}$ (anti)neutrinos of all flavors
with typical energies of 10--15~MeV. Given the large amount of
neutrinos involved, it is expected that they will play an important
role in both the supernova dynamics and the underlying
nucleosynthesis.

At present, the delayed neutrino-heating
mechanism~\cite{Bethe.Wilson:1985} represents the most promising
scenario to explain supernova explosions. In fact, recent
two-dimensional
simulations~\cite{Mueller.Janka.Marek:2012,Bruenn.Mezzacappa.ea:2013,Suwa.Takiwaki.ea:2013}
produce neutrino driven explosions on time scales of several hundreds
of milliseconds after bounce. Once the explosion sets in, the
continuous emission of neutrinos from the PNS drives a low-mass
outflow known as neutrino-driven
wind~\cite{duncan.shapiro.wasserman:1986} that is currently considered
a favored site for the productions of elements heavier than iron
(e.g.~\cite{Arcones.Thielemann:2013}). As neutrinos travel through the
stellar mantle, they can suffer flavor
oscillations~\cite{Duan.Kneller:2009,Duan.Fuller.Qian:2010},
contribute to the nucleosynthesis of several rare
isotopes~\cite{Woosley.Hartmann.ea:1990,Heger.Kolbe.ea:2005} and even
drive an r process in the He-shell of metal-poor
stars~\cite{Banerjee.Haxton.Qian:2011} before they are finally
detected on Earth.

This paper addresses several issues related to the role of neutrinos
on the nucleosynthesis of heavy elements in core-collapse supernova
explosions. After discussing recent constraints from metal-poor stars
observations on the production of heavy elements by the r-process
(section~\ref{sec:metal-poor-star}), we describe different nuclear
physics aspects that are relevant to determine the luminosities and
spectra of neutrinos emitted during the PNS deleptonization which
proceeds via weak processes. For those involving nucleons
(section~\ref{sec:charg-curr-neutr}), corrections due to the nuclear
interaction are relevant due to the relatively high density in the
decoupling region. They are particularly important for charged-current
processes and are related to the symmetry energy of nuclear
matter. New long term Boltzmann transport simulations of the
neutrino-driven wind are presented in
section~\ref{sec:nucl-neutr-winds} and its nucleosynthesis yield
discussed.

\section{Implications from metal-poor star observations}
\label{sec:metal-poor-star}

Traditionally the production of elements heavier than iron has been
associated with two neutron capture processes, the r-process and the
s-process~\cite{Burbidge.Burbidge.ea:1957,Cameron:1957} with a smaller
contribution due to the p-process (or
$\gamma$-process)~\cite{Arnould.Goriely:2003}. In this picture, the
abundances of heavy elements observed in our Sun are the superposition
of different events producing s-process elements (intermediate-mass
AGB stars and massive
stars)~\cite{straniero.gallino.cristallo:2006,Busso.Gallino.Wasserburg:1999,Pignatari.Gallino.ea:2010}
and r-process elements in a so far not yet identified
site(s)~\cite{arnould.goriely.takahashi:2007}. The so-called solar
system r-process abundance is obtained by sustracting from the solar
system isotopic
abundances~\cite{Lodders:2003,Asplund.Grevesse.ea:2009,Jacobson.Frebel:2013}
the solar s-process pattern determined from a combination of
experimentally determined $(n,\gamma)$ cross sections and AGB stellar
models~\cite{Arlandini.Kaeppeler.ea:1999,Bisterzo.Gallino.ea:2010}.

Due to the long evolutionary time scales of intermediate mass stars
before reaching the AGB phase (up to a billion years), it is expected
that stars that are sufficiently old will have been enriched by the
r-process but not by the s-process. (Provided that the r-process
operates in a site with a much shorter evolutionary time scale than
the s-process sites). Astronomers use the metallicity, normally
associated with the abundance of Iron in the star, as a proxy for the
age. Metal-poor stars, $[\mathrm{Fe/H}] \lesssim
-1.5$\footnote{Astronomers use the [A/B] notation to describe the
  relative abundances of two elements in a star compared to that in
  the Sun: $[\text{A/B}] = \log_{10} (N_{\text{A}}/N_{\text{B}})_*
  -\log_{10} (N_{\text{A}}/N_{\text{B}})_\odot$}, present abundances
of heavy elements that do not correspond with the solar system
s-process
abundances~\cite{Cowan.Sneden:2006,Sneden.Cowan.Gallino:2008}. These
stars, also called r-II stars~\cite{Christlieb.Beers.ea:2004}, present
robust abundance patterns that match almost perfectly the solar system
r-process abundance pattern for elements with
$Z>50$~\cite{Cowan.Sneden:2006,Sneden.Cowan.Gallino:2008,Roederer.Lawler:2012}. Some
other stars~\cite{Honda.Aoki.ea:2006,Roederer.Lawler.ea:2012} are
depleted in elements heavier than $Z=50$ but enriched in lighter
elements with $A\sim 90$ like Sr, Y and Zr.  As these elements are
abundantly produced in alpha-rich freeze-outs from
neutrino-driven winds~\cite{Woosley.Hoffman:1992,Witti.Janka.Takahashi:1994}
(see also section~\ref{sec:nucl-neutr-winds}) by charged-particle
reactions (CPR) they are sometimes denoted CPR
elements~\cite{Qian.Wasserburg:2007}.  This reference suggested a
simple ``LEGO-block'' model to explain these observations. In this
model the production of CPR (also known as light r-process) elements
with $Z\lesssim 50$ and heavy r-process elements with $Z>50$ is due to
two different sources: 1) an H-source produces the heavy r-process
elements together with CPR elements and no Iron with an abundance
pattern as given by r-II stars; 2) an L-source produces light
r-process elements or CRP elements with an abundance pattern given by
the observations of
HD~122563~\cite{Honda.Aoki.ea:2006,Roederer.Lawler.ea:2012} together
with Iron. In this model the H-source is the site where the ``main''
r-process operates while the L-source produces light r-process
elements in what sometimes is denoted as ``weak'' r-process. Similar
results were obtained in the Galactic chemical evolution study of
ref.~\cite{Travaglio.Gallino.ea:2004} that suggested the existence of
a ``lighter element primary process'' (LEPP) to explain the abundances
of Sr, Y and Zr at low metallicities. Here ``process'' has to be
interpreted as a different astrophysical site to the one in which the
``main'' r-process operates (in this case the LEPP is analogous to
the L-source of ref.~\cite{Qian.Wasserburg:2007}).

\begin{figure}[htb]
  \centering
  \includegraphics[width=\linewidth]{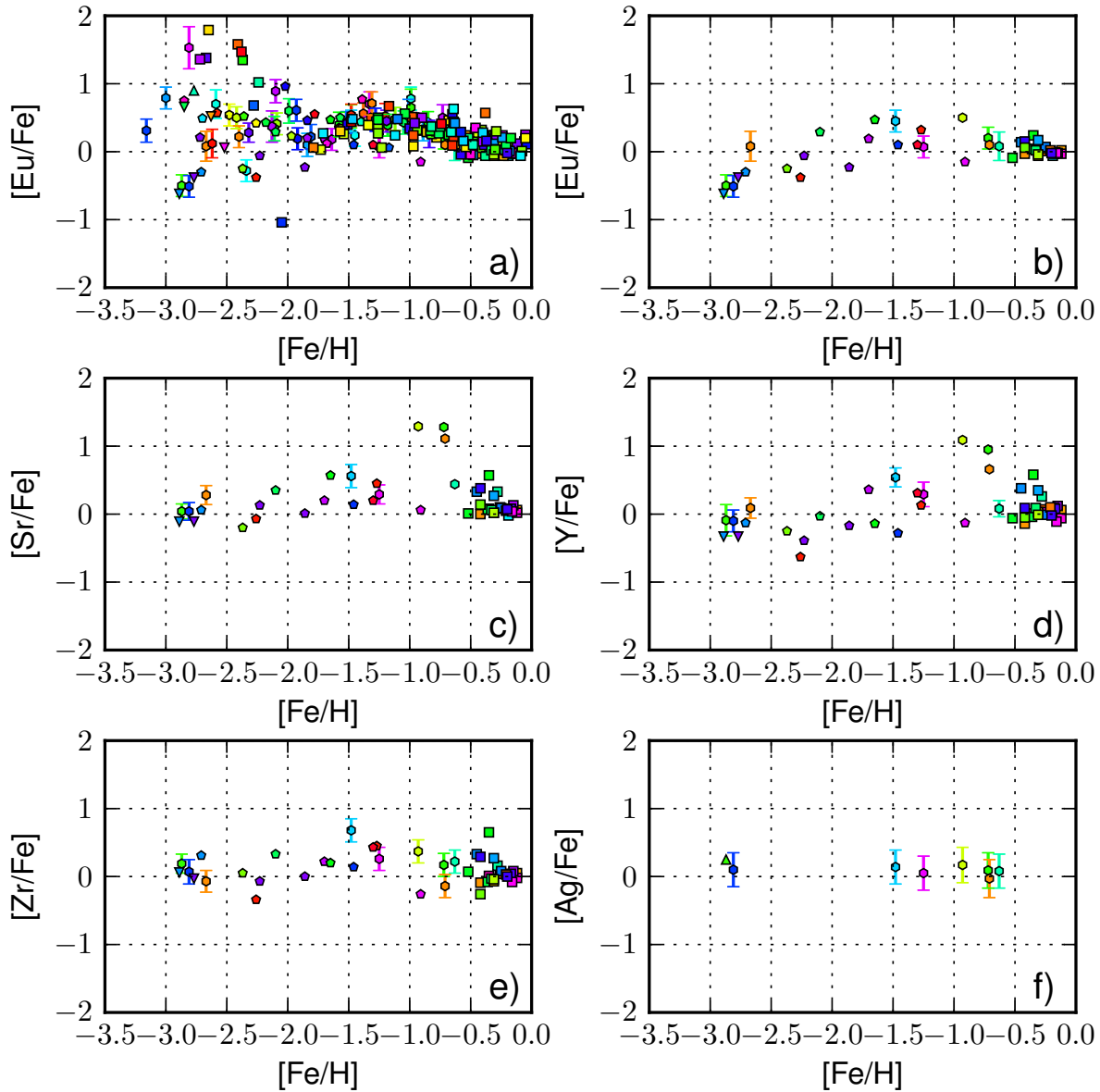}
  \caption{Abundance ratio of metal-poor
    stars~\cite{Burris.Pilachowski.ea:2000,Honda.Aoki.ea:2004,Barklem.Christlieb.ea:2005,Hansen.Primas.ea:2012}
    as a function of metallicity, [Fe/H]. The upper left panel
    includes all metal-poor stars in the sample while all other panels
    include only stars whose abundance pattern can be related to an
    L-source event. They are enriched in light r-process elements and
    poor in heavy r-process elements.\label{fig:abundevol}}
\end{figure}

The above model explains the large scatter in the abundance of Eu
observed at low
metalicities~\cite{Cowan.Thielemann:2004,Sneden.Cowan.Gallino:2008}
(see upper left panel of figure~\ref{fig:abundevol}). At early times
the Galaxy is chemically unmixed and Eu is mainly produced in rare
events with little or no iron
production~\cite{Wasserburg.Qian:2000}. This model predicts that there
may be stars that are only enriched in light r-process elements but so
far all metal-poor stars observed contain both light and heavy
r-process elements~\cite{Roederer:2013}. It may indicate that the
relative low abundances of heavy r-process elements observed in stars
that show an abundance pattern like HD~122563 are in fact produced
together with light r-process elements but with much lower
efficiency. If the L-source site is neutrino-driven winds from
core-collapse supernova as suggested by
ref.~\cite{Qian.Wasserburg:2007} it implies that neutrino-driven winds
should produce also elements with $Z>50$. In principle, there could be
a large variability in the production of light r-process elements from
event to event, however the work of ref.~\cite{Montes.Beers.ea:2007}
suggest that their production is robust and always results in an
abundance pattern similar to the one observed in HD~122563.

If Eu ($Z=63$) is in fact produced together with Fe in L-source events
a plot showing stars that have been enriched by these events should
show a correlation, i.e. [Eu/Fe] should be constant as a function of
metallicity. A star whose abundance pattern is due to an H-source
event has a ratio Eu/Sr that is much larger than the one found in the
solar system. Typical values are $[\text{Eu/Sr}] \sim 1$. Hence, we
characterize stars polluted by an L-source event, i.e. enriched in
light r-process elements, as those with $[\text{Eu/Sr}] < 0$. A lower
value could also be used but in this case the number of stars in the
sample is largely reduced. This condition includes also stars that
have been enriched mainly by the s-process but they should not be
present at low metallicities. Figure~\ref{fig:abundevol} shows the
[Eu/Fe], [Sr/Fe], [Y/Fe], [Zr/Fe], and [Ag/Fe] for a sample of
metal-poor stars from recent
compilations~\cite{Burris.Pilachowski.ea:2000,Honda.Aoki.ea:2004,Barklem.Christlieb.ea:2005,Hansen.Primas.ea:2012}. Only
stars with $[\text{Eu/Sr}] < 0$ are shown. It is intriguing to see
that a correlation in [Eu/Fe] versus metallicity emerges that is
absent once the whole sample of metal-poor stars is considered. This
may suggest that Eu is co-produced together with Fe in the events that
produce light r-process elements. A correlation is also found in Sr, Y
and Zr that is expected if these elements are produced in
neutrino-driven winds~\cite{Qian.Wasserburg:2007}. The scatter is also
greatly reduced when compared to a sample without any selection
criteria (see figures in
refs.~\cite{Travaglio.Gallino.ea:2004,Hansen.Primas.ea:2012}). The
situation is unclear for Ag as only two stars have survived our
criteria. A larger sample of metal-poor stars is required to check
this hypothesis.  Ref.~\cite{Hansen.Primas.ea:2012} has found an
anticorrelation between Ag and Pd and Sr, Y and Zr suggesting that
both element groups are produced in different sites.

In the above discussion, we have explicitly avoided to mention
possible sites for r-process nucleosynthesis. Several sites have been
suggested for both the H-source and L-source events. It is likely that
the production of both light and heavy r-process elements is dominated
by different sites as a function of metallicity. Neutron star mergers
are a likely site for H-source events. Recent
simulations~\cite{Korobkin.Rosswog.ea:2012,Bauswein.Goriely.Janka:2013}
show that the material ejected dynamically during the merger is an
ideal site for a robust r-process that produces nuclei with $A \gtrsim
120$ ($Z \gtrsim 50$) in agreement with the solar r-process abundance
distribution~\cite{Korobkin.Rosswog.ea:2012,Bauswein.Goriely.Janka:2013}. In
this scenario, the production of nuclei with $A\sim 120$ is due to the
fission of nuclei in the region of $A\sim 280$ and sensitive to the
particular fission rates and yields
used~\cite{Martinez-Pinedo:2008,Petermann.Langanke.ea:2012}.  Fission
cycling is also responsible for producing a robust r-process
pattern. An important aspect of r-process in neutron star mergers is
that the amount of r-process material ejected is large enough
($10^{-3}$~M$_\odot$ to $10^{-2}$~M$_\odot$) to produce an
optical/near-infrared transient powered by the radioactive decay of
r-process
nuclei~\cite{Metzger.Martinez-Pinedo.ea:2010,Roberts.Kasen.ea:2011,Bauswein.Goriely.Janka:2013}. The
transient luminosity is expected to reach a maximum of $\sim
10^{42}$~erg~s$^{-1}$, i.e. kilonova luminosities, in timescales of
hours to
days~\cite{Metzger.Martinez-Pinedo.ea:2010,Roberts.Kasen.ea:2011,Bauswein.Goriely.Janka:2013}. The
observation of a near-infrared transient associated with the short
$\gamma$-ray burst GRB130603B has been related to a kilonova type
event powered by the decay of r-process
nuclei~\cite{Tanvir.Levan.ea:2013,Berger.Fong.Chornock:2013} (however
see the caveats mentioned in
ref.~\cite{Grossman.Korobkin.ea:2013}). If confirmed, this will be the
first direct observation of an r-process nucleosynthesis event and
will demonstrate that neutron-star mergers are an r-process
site. Neutron-star mergers are also expected to eject material from
outflows from the accretion disk formed around the black hole
remnant. This material will be ejected either by
neutrino-driven winds~\cite{Wanajo.Janka:2012,Grossman.Korobkin.ea:2013} or
by viscous heating and recombination of nucleons into alpha
particles~\cite{Fernandez.Metzger:2013}. These ejecta are expected to
produce light r-process elements making neutron-star mergers a
possible candidate to account for observations of r-II metal-poor
stars. However, due to the long evolutionary timescales involved,
neutron-star mergers have problems to explain for the r-process at low
metallicities~\cite{Qian:2000,Argast.Samland.ea:2004}. Several sites
have been suggested that are expected to operate mainly at low
metalicities: Jets from magnetorotational
supernova~\cite{Winteler.Kaeppeli.ea:2012} and a neutrino-induced
r-process in the He shell of core-collapse
supernova~\cite{Banerjee.Haxton.Qian:2011}. These sites produce mainly
nuclei with $A\gtrsim 130$ but an additional contribution producing
light r-process elements from neutrino-driven winds from the protoneutron
star or black-hole accretion disk is expected.

Neutrino-driven winds from protoneutron stars have been suggested as the site
for L-source events~\cite{Qian.Wasserburg:2007}, i.e. light r-process
elements. This aspect is analyzed in
section~\ref{sec:nucl-neutr-winds} based on Boltzmann neutrino
transport simulations. 

\section{Charged-current neutrino interactions in core-collapse
  supernovae} 
\label{sec:charg-curr-neutr}

Neutrino-driven winds from protoneutron stars have been considered for a long
time like a possible site for the nucleosynthesis of heavy elements
(see ref.~\cite{Arcones.Thielemann:2013} for a recent review). It is
expected that independently of the explosion mechanism a protoneutron
star (PNS) will form after a core-collapse supernova explosion. The
PNS deleptonizes by continuous emission of neutrinos of all flavors in a
period of several tens of seconds.  Neutrino absorption processes at
the PNS surface deposit heat which drives a matter outflow known as
the neutrino-driven wind~\cite{duncan.shapiro.wasserman:1986}. Initial
studies showed that neutrino-driven winds could be the site for the
r-process~\cite{Woosley.Wilson.ea:1994,Takahashi.Witti.Janka:1994}. These
pioneering works where followed by analytic~\cite{Qian.Woosley:1996},
parametric~\cite{hoffman.woosley.qian:1997} and steady state wind
models~\cite{Otsuki.Tagoshi.ea:2000,Thompson.Burrows.Meyer:2001} that
showed that neutrino-driven winds produce both light and heavy r-process
elements provided that the outflow has short dynamical time scales (a
few milliseconds), high entropies (above 150~$k$/nucleon) and low
electron fractions ($Y_e < 0.5$). Recent hydrodynamical
simulations~\cite{Arcones.Janka.Scheck:2007,Arcones.Janka:2011} show
that the short dynamical timescales can in fact be achieved but fail
however to obtain the necessary entropies at times relevant for
r-process nucleosynthesis~\cite{Kuroda.Wanajo.Nomoto:2008}. These
works rule out the posibility that neutrino-driven winds are
responsible for the production of heavy r-process elements with
$Z\gtrsim 50$. In the following, we analyze the possibility that light
r-process elements ($Z\gtrsim 50$) may be produced in neutrino winds.

The nucleosynthesis outcome of neutrino-driven winds is very sensitive
to the electron fraction, $Y_e$, of the ejected
matter~\cite{Roberts.Woosley.Hoffman:2010,Arcones.Montes:2011} that is
determined by the competition between electron neutrino absorption in
neutrons and antineutrino absorption in neutrons and their inverse
reactions. These rates are rather sensitive to the luminosity and
spectral differences between electron neutrinos and antineutrinos. The
work of ref.~\cite{Arcones.Montes:2011} has shown that light r-process
elements can be produced both in proton-rich and neutron-rich
ejecta. However, the assumed $Y_e$ evolutions need to be reevaluated
on the light of recent
works~\cite{Martinez-Pinedo.Fischer.ea:2012,Roberts:2012,Roberts.Reddy.Shen:2012}. This
aspect has recently been revisited in ref.~\cite{Wanajo:2013}.

Deep in the interior of the protoneutron star neutrinos are in thermal
and chemical equilibrium with matter. However, as we move to the
surface and the temperature and density drop neutrinos decouple with
matter. As $\mu$ and $\tau$ neutrinos interact only via neutral
current they are the first to decouple. For the very neutron rich
conditions found at the PNS surface, electron antineutrinos decouple
before electron neutrinos. As the neutrino spectrum reflects the local
properties of matter at the position in which they decouple, one
expects  the following hierarchy of neutrino energies:
$\varepsilon_{\nu_{\mu,\tau}} > \varepsilon_{\bar{\nu}_e} >
\varepsilon_{\nu_e}$~\cite{keil.raffelt.janka:2003}, with $\varepsilon
= \langle E^2\rangle/\langle E \rangle$. The fact that electron
antineutrinos have larger average energies than electron neutrinos
suggest neutron rich ejecta. However, one has also to
consider the other energy scale in the problem, i.e. the neutron to
proton mass difference. It turns out that neutron-rich ejecta are only
obtained when $\varepsilon_{\bar{\nu}_e} - \varepsilon_{\nu_e} > 4
(m_n -
m_p)$~\cite{Qian.Woosley:1996,Froehlich.Hauser.ea:2006,Martinez-Pinedo:2008b}. As
the treatment of neutrino transport and neutrino matter interactions
improved over the years the energy difference between electron
antineutrinos and neutrinos decreased and different
simulations~\cite{Buras.Rampp.ea:2006,Liebendoerfer.Mezzacappa.ea:2001a,Thompson.Quataert.Burrows:2005}
obtained proton-rich ejecta during the early times of the
explosion. Supernova simulations based on three-flavor Boltzmann
neutrino transport have been extended to time scales of several tens
of
seconds~\cite{Huedepohl.Mueller.ea:2010,Fischer.Whitehouse.ea:2010},
covering the whole deleptonization of the PNS. They have predicted a
continuous decrease of the energy difference between neutrinos and
antineutrinos of all flavors that became practically indistinguishable
after $\approx 10$~s.

With the development of three-flavor Boltzmann neutrino transport
codes it has been possible to relate the spectra of the emitted
neutrinos and the underlying nucleosynthesis to high density neutrino
matter interactions and basic properties of the nuclear equation of
state. Recently, it has been shown that the energy difference between
electron neutrinos and antineutrinos, and consequently the $Y_e$ of
the ejecta, is very sensitive to the treatment of charged-current
reactions $\nu_e + n \rightarrow p + e^-$ and $\bar{\nu}_e +p
\rightarrow n + e^+$ in neutron rich matter at densities $\rho \approx
10^{12}$--$10^{14}$~g~cm$^{-3}$~\cite{Martinez-Pinedo.Fischer.ea:2012,Roberts:2012,Roberts.Reddy.Shen:2012}. 

\begin{figure}[htb]
  \centering
  \includegraphics[width=\linewidth]{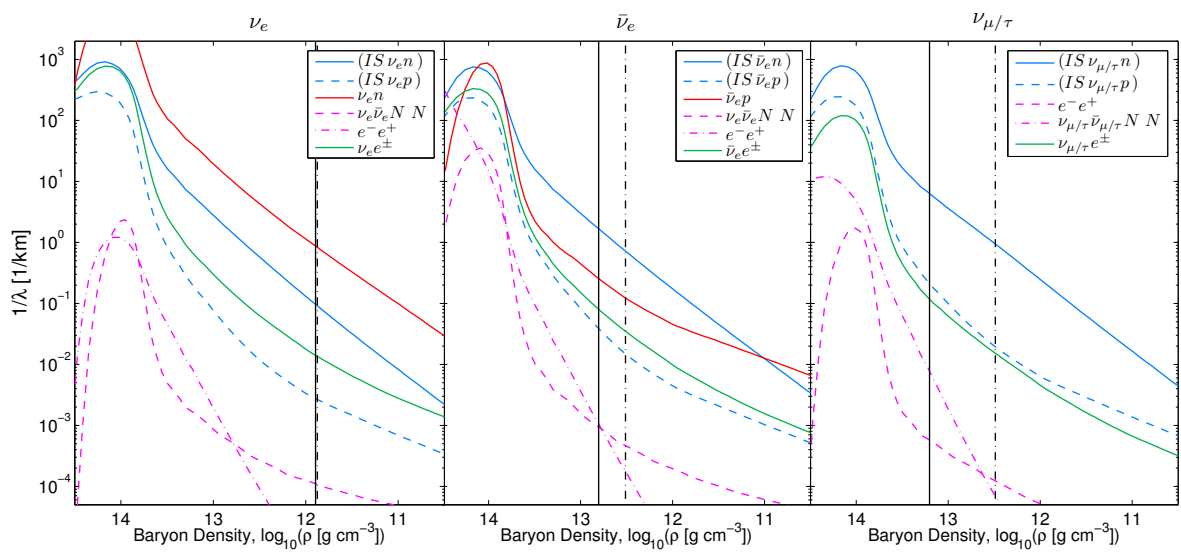}
  \caption{Inverse mean free paths for the main reactions contributing
    to the determination of the spectra of neutrinos emitted from a
    protoneutron star: isoenergetic neutrino-nucleon scattering
    $(\text{IS}, \nu N)$, charged-current reactions ($\nu_en$,
    $\bar\nu_ep$), $N$--$N$--Bremsstrahlung ($\nu\bar\nu N N$),
    $e^-e+$ annihilation, and neutrino electron/positron scattering
    ($\nu e^{\pm}$). The different panels show the inverse mean free
    path for $\nu_e$ (left), $\bar{\nu}_e$ (middle) and $\nu_{\mu,\tau}$
    (right) based on radial profiles at 1~second post bounce for the
    18.0~M$_\odot$ progenitor simulations of
    ref~\cite{Fischer.Whitehouse.ea:2010}. The vertical black solid
    and dash-dotted lines mark the position of the energy and
    transport neutrinospheres. Figure adapted from
    ref.~\cite{Fischer.Martinez-Pinedo.ea:2012}. \label{fig:opacities}}
\end{figure}

Figure~\ref{fig:opacities} shows the opacities or inverse mean free
paths for the main processes determining the interaction of
(anti)neutrinos with matter. The figure shows also the position of the
energy and transport neutrinospheres for the different neutrino
flavors (see ref.~\cite{Fischer.Martinez-Pinedo.ea:2012} for a
description of the different processes and the determination of the
neutrinospheres). One clearly sees the hierarchy mentioned above with
$\nu_{\mu,\tau}$ (anti)neutrinos decoupling at the highest densities,
$\bar{\nu}_e$ at intermediate densities and $\nu_e$ at the lowest
densities. For the conditions considered in the figure (around 1~s
postbounce) the neutrinospheres are located in the density range
$10^{12}$--$10^{13}$~g~cm$^{-3}$, i.e. $10^{-3}$--$10^{-2}$~$\rho_0$
with $\rho_0$ the nuclear saturation density. Only for times around
10~s or later do the neutrinospheres move at densities around the
saturation density. 

Neutrinos are in thermal equilibrium with matter for densities larger
than the location of the energy sphere, at lower densities neutrinos
have decoupled from matter but still suffer several scattering events
with nucleons till they reach the transport
neutrinosphere~\cite{Raffelt:2001}. The region between the energy and
transport spheres constitutes a scattering atmosphere where due to
the slightly inelastic nature of neutrino-nucleon scattering the spectra
of neutrinos is modified~\cite{Raffelt:2001}. This aspect is important
for the determination of the $\bar{\nu}_e$ and $\nu_{\mu,\tau}$
spectra (see figure~\ref{fig:opacities}).

The thermal equilibrium of neutrinos with matter is maintained by
processes in which the neutrino exchanges energy. As can be seen
from~\ref{fig:opacities}, the most important 
processes are charged-current $\nu_e + n \rightarrow p + e^-$ and
$\bar{\nu}_e +p \rightarrow n + e^+$. For the process $\nu_e + n
\rightarrow p + e^-$, the opacity or inverse mean free path of a
neutrino of energy $E_\nu$ is given
by~\cite{Reddy.Prakash.Lattimer:1998,Roberts.Reddy.Shen:2012}:

\begin{equation}
  \label{eq:1}
  \chi(E_\nu) = \frac{G_F^2 V_{ud}^2}{(\hbar c)^7} (g_V^2+3 g_A^2) \int
  \frac{d^3p_e}{(2\pi)^3} [1-f_e(E_e)] S(q_0,q) 
\end{equation}
where $f$ is the Fermi-Dirac distribution function and $S(q_0,q)$ is
the response function of the nuclear medium that is assumed identical
for Fermi and Gamow-Teller operators. The energy transfer to the
medium is $q_0 = E_\nu - E_e$ and the magnitude of the momentum
transfer is $q=|\bm{p}_\nu - \bm{p}_e|$. Equations of State commonly
used in core-collapse supernova simulations (see, e.g.,
refs~\cite{Lattimer.Swesty:1991,Shen.Toki.ea:1998a}) treat neutrons
and protons as a gas of non-interacting quasiparticles that move in a
mean-field single-particle potential, $U$. In relativistic mean-field
models the energy-momentum dispersion relation becomes:

\begin{equation}
  \label{eq:2}
  E_i(p) = \sqrt{p^2 + m_i^{*2}}+U_i, 
\end{equation}
where $m_i^{*}$ is the nucleon effective mass. For the neutron-rich
conditions around the neutrinospheres the mean-field potential for
neutrons and protons can be very different with their relative
difference $\Delta U = U_n - U_p$ directly related to the nuclear
symmetry
energy~\cite{Reddy.Prakash.Lattimer:1998,Haensel.Potekhin.Yakovlev:2007}. Ref.~\cite{Reddy.Prakash.Lattimer:1998}
provides analytic expressions for the response function valid both in
the relativistic and non-relatistic limits that can be used for the
calculation of the opacity. A simplified expression can be obtained
assuming that zero momentum transfer, $q \approx 0$, reflecting the
fact that nucleons are more massive than leptons. In this case, the
energy transfer becomes $q_0 = -(m^*_n - m^*_p) - \Delta U = -\Delta
m^* -\Delta U$ and the opacity is given by (elastic
approximation)~\cite{Reddy.Prakash.Lattimer:1998,Bruenn:1985}:

\begin{equation}
  \label{eq:3}
  \chi(E_{\nu}) = \frac{G_F^2 V_{ud}^2}{\pi (\hbar c)^4} (g_V^2+3
  g_A^2) p_e E_e [1-f_e(E_{e})] \frac{n_n - n_p}{1-e^{(\varphi_p - \varphi_n)/T}}, 
\end{equation}
with $\varphi_i = \mu_i -m^*_i-U_i$, $\mu_i$ the chemical
potential of the nucleon and $E_e = E_\nu + \Delta m^* + \Delta U$. If
we consider neutrinos with energies smaller than $E_\nu^0 = \mu_e
- \Delta m^* - \Delta U$, that varies between  10 and 30~MeV when the
density varies in the range $10^{12}$--$10^{13}$~g~cm$^{-3}$, the
opacity behaves as:

\begin{equation}
  \label{eq:4}
  \chi(E_\nu) \propto (E_\nu + \Delta m^* + \Delta U)^2
  \exp\left(\frac{E_\nu + \Delta m^* + \Delta U - \mu_e}{T}\right).  
\end{equation}

The opacity increases by an exponential factor due to the inclusion of
mean-field modifications, $\Delta U$, to the vacuum Q-value, $\Delta
m^*$. Due to the strong sensitivity of the opacity to the neutrino
energy, the density at which neutrinos decouple increases with
decreasing neutrino energy. However, $\Delta U$ increases with density
implying that the smaller is the energy of the neutrino emitted from
the protoneutron star the larger the opacity correction due to the
inclusion of mean-field effects (see figure~6 of
ref.~\cite{Roberts.Reddy.Shen:2012}) . As the opacity for neutrinos
mainly determines the deleptonization rate, i.e. the neutrino
luminosity of the protoneutron star, we expect that the larger the
$\Delta U$ correction, i.e. the larger the symmetry energy, the
smaller the neutrino luminosity. This is in fact confirmed by
Boltzmann transport
simulations~\cite{Martinez-Pinedo.Fischer.ea:2012,Roberts.Shen.ea:2012}.

For $\bar{\nu}_e$  there is no final state blocking for the produced
positron and consequently the main effect of mean-field corrections is
to change the energy threshold for neutrino absorption. Using a similar
analysis than the above it can be shown that the opacity for
antineutrinos behaves like:

\begin{equation}
  \label{eq:5}
  \chi(E_{\bar{\nu}_e}) \propto (E_{\bar{\nu}_e} - \Delta m^* - \Delta U)^2.
\end{equation}

The discussion above shows that mean-field effects increase the
opacity for neutrino absorption while reducing the opacity for
antineutrino absorption. When compared with simulations that do not
include mean-field effects the increase of opacity for neutrinos will
keep them in thermal equilibrium with matter up to larger radii. They
decouple in regions of lower temperature and consequently their
average energy is smaller. The average energy of antineutrinos is
expected to increase as they decouple at slightly deeper (hotter)
regions of the neutron star. However, due to the reduced dominance of
charged-current reactions (see figure~\ref{fig:opacities}) the change
in average energy is expected to be smaller for $\bar{\nu}_e$. This
has been confirmed by recent long term simulations of PNS
cooling~\cite{Martinez-Pinedo.Fischer.ea:2012,Roberts:2012,Roberts.Reddy.Shen:2012}
that treat charged-current opacities consistently with the EoS at the
mean-field level. They have shown that the inclusion of mean-field
effects increases the energy difference between $\bar{\nu}_e$ and
$\nu_e$ with respect to the values obtained without mean-field
effects. For the nucleosynthesis point view, the main consequence is
that ejecta that were proton-rich during the whole deleptonization
period of the protoneutron star now become neutron-rich during the
first seconds and later turn proton-rich as the energy difference
between $\bar{\nu}_e$ and $\nu_e$ decreases. The neutron-richness of
the ejecta depends on the symmetry energy of the underlying
EoS. Hence, it becomes important to use an EoS that reproduce recent
constraints on the symmetry energy.

Notice that the symmetry energy does not only affect the evolution
during the deleptonization phase but determines the whole structure of
the protoneutron star after the onset of the explosion. Electron
captures on nuclei determine the amount of deleptonization occurring
during the collapse~\cite{Langanke.Martinez-Pinedo:2003}. At the onset
of neutrino trapping, at densities around $10^{12}$~g~cm$^{-3}$, $Y_e$
reaches a value $Y_e^{\text{trap}} \approx 0.3$~\cite{martinez-pinedo.liebendoerfer.frekers:2006}. At this moment
electron (anti)neutrinos thermalize and the system reaches weak
equilibrium, $\mu_{\nu_e} + \mu_n = \mu_e + \mu_p$. Once the core
reaches nuclear matter densities and a transition to uniform nuclear
matter occurs the relative ratio of neutrons and protons (and electrons
and neutrinos) is fully determined by the symmetry energy under the
constrain of constant lepton number, $Y_{\text{lep}} = Y_e + Y_{\nu_e}
= Y_e^{\text{trap}}$. This can be seen from the following
relation~\cite{Haensel.Potekhin.Yakovlev:2007}: 

\begin{equation}
  \label{eq:6}
  \mu_e - \mu_{\nu_e} = \mu_n - \mu_p = 4(1-2 Y_e) S(\rho),
\end{equation}
with $S(\rho)$ the nuclear symmetry energy. As protons and neutrinos
have rather low abundances, their exact value is rather sensitive to
changes of the symmetry energy. This implies that in order to
determine the impact of a particular EoS on the nucleosynthesis
outcome of neutrino-driven winds it is important to cover all relevant
supernova phases including core-collapse, bounce, post-bounce
accretion, explosion and cooling. In particular, the high $Y_e$ values
obtained at early times ($\lesssim 1$~s) in the protoneutron star
evolution simulations of ref.~\cite{Roberts:2012} could be an artifact
of the particular progenitor used. This caveat was noted in
ref.~\cite{Roberts:2012} but was not accounted for in the
nucleosynthesis study of ref.~\cite{Wanajo:2013}. Due to their large
mass loss rate~\cite{Wanajo:2013}, the early cooling phase is
particularly important for determining the mass-integrated
nucleosynthesis of neutrino-driven winds. To accurately resolve the
transition between accretion and cooling phases  requires
multidimensional simulations~\cite{Mueller.Janka.Heger:2012}.

\begin{figure}[htb]
  \centering
  \includegraphics[width=0.5\linewidth]{eos.eps}%
  \includegraphics[width=0.5\linewidth]{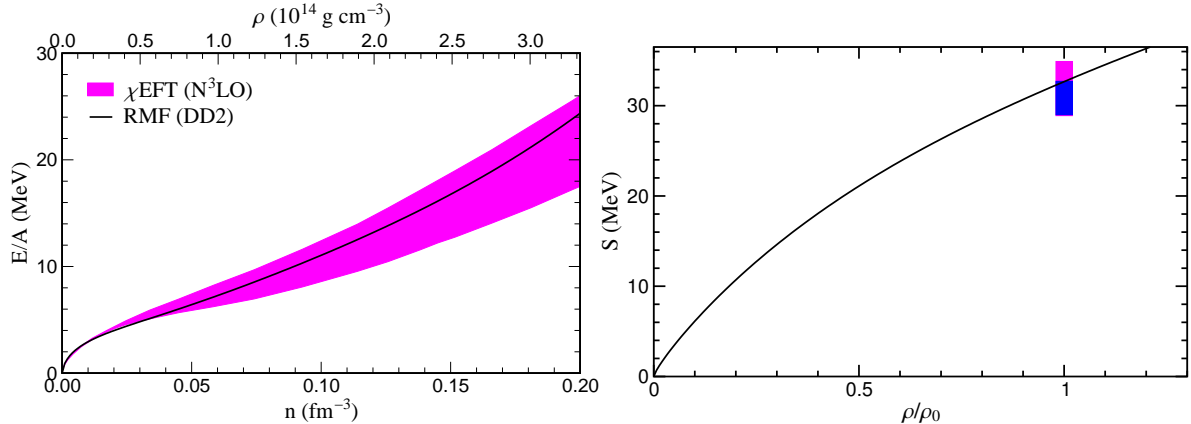}%
  \caption{(left panel) Energy per nucleon for neutron matter
    computed by chiral EFT interactions computed up to N$^3$LO
    order~\cite{Tews.Krueger.ea:2013,Krueger.Tews.ea:2013}. The
    magenta band includes uncertainty estimates due to the many-body
    calculation, low-energy constants and the regularization cutoffs
    in the 3N and 4N forces (see ref.~\cite{Krueger.Tews.ea:2013} for
    additional details). The black line shows the energy per nucleon
    predicted by RMF calculations using the density dependent DD2
    functional. (right panel) Symmetry energy, defined as the
    difference between the energy per nucleon for neutron and
    symmetric matter, determined by the RMF DD2 calculations compared
    with recent constrains from chiral EFT~\cite{Krueger.Tews.ea:2013}
    (magenta box) and nuclear physics experiments and astronomical
    observations~\cite{Lattimer.Lim:2013} (blue box). \label{fig:eos}}
\end{figure}

The equation of state for neutron-matter has been recently computed
using all many-body forces among neutrons predicted by chiral
effective field theory (EFT) up to next-to-next-to-next-to-leading
order (N$^3$LO)~\cite{Tews.Krueger.ea:2013,Krueger.Tews.ea:2013}. The
predicted energy per nucleon for neutron-matter is shown in the left
panel of figure~\ref{fig:eos}. When compared with equations of state
used in core-collapse simulations (see figure~9 of
ref.~\cite{Krueger.Tews.ea:2013}) one finds that the two most commonly
used, the non-relativistic Skyrme-like of Lattimer and
Swesty~\cite{Lattimer.Swesty:1991} and the relativistic mean-field
(RMF) based on the TM1 functional~\cite{Shen.Toki.ea:1998a}, are not
consistent with chiral EFT constraints particularly at subsaturation
densities. An EoS that reproduces these constraints is the RMF based
on the density-dependent DD2 functional~\cite{Typel.Roepke.ea:2010}
(see figure~\ref{fig:eos}). In addition, the DD2 based EoS reproduces
constraints on the symmetry energy at saturation density from chiral
EFT~\cite{Krueger.Tews.ea:2013} and a global analysis combining
nuclear experimental information and astronomical observations of
neutron stars~\cite{Lattimer.Lim:2013}. The symmetry energy predicted
by the DD2 EoS is shown in the right panel of fig.~\ref{fig:eos} for
the whole range of densities relevant for neutrino-wind
simulations. Notice that clustering effects that are expected to be
important at subsaturation densities~\cite{horowitz.schwenk:2006} are
not included in the calculation of the symmetry energy. The DD2 EoS
also reproduces constraints on the maximum mass of neutron stars and
comes close (within 1~km) to the recent estimates of the observational
radius~\cite{Fischer.Hempel.ea:2013}.

\section{Nucleosynthesis in neutrino-driven winds}
\label{sec:nucl-neutr-winds}

\begin{figure}[htb]
  \centering
  \includegraphics[width=\linewidth]{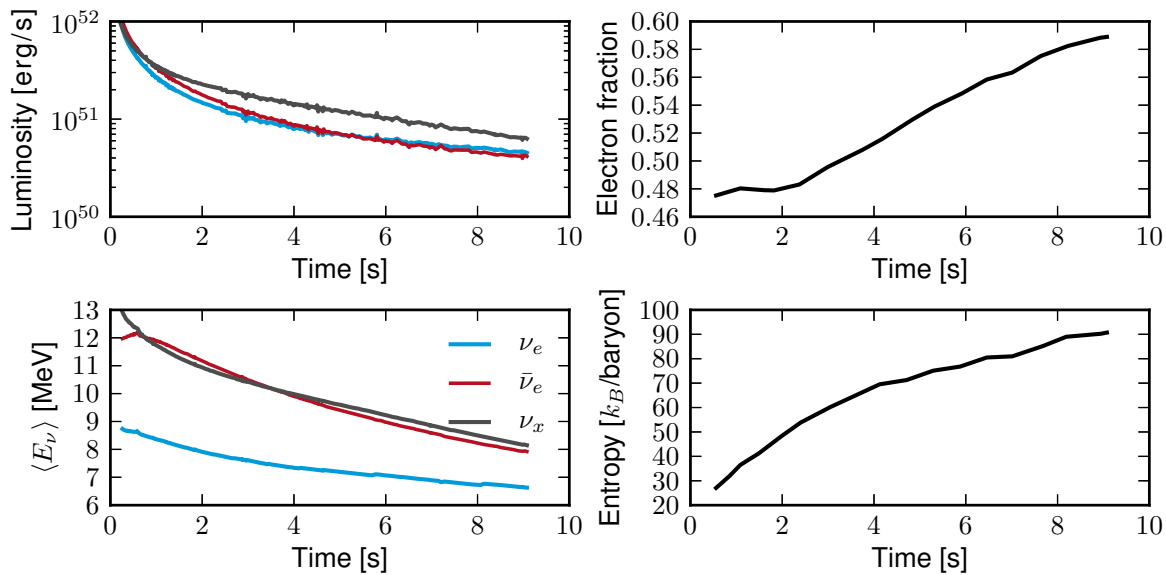}
  \caption{The left panels show the evolution of neutrino
    luminosities (upper) and average neutrino energies (lower) for the
    different neutrino flavors. The right panels show the
    asymptotic values of entropy (lower) and $Y_e$ (upper) reached in
    the ejecta. \label{fig:wind11}}
\end{figure}

We have performed core-collapse supernova simulations based on
spherically symmetric radiation hydrodynamics with three-flavor
Boltzmann neutrino transport using the AGILE-Boltztran code. In the
high density regime, we use the DD2 EoS table provided by
M. Hempel\footnote{\url{http://phys-merger.physik.unibas.ch/~hempel/eos.html}}. The
equation of state is based on the extended nuclear statistical model
of ref.~\cite{Hempel.Schaffner-Bielich:2010} and includes a detailed
nuclear composition allowing for the presence of light nuclear
clusters at subsaturation densities. Importantly, the EoS provides the
mean-field corrections necessary for the calculation of
charged-current neutrino reactions. Additional details will be
provided in a forthcoming publication. The simulations are based on
the 11.2~M$_\odot$ progenitor of
ref.~\cite{Woosley.Heger.Weaver:2002}.  Because spherically symmetric
simulations do not result in explosions for this iron-core progenitor,
we enhance the neutrino heating rates in the gain region following the
scheme of ref.~\cite{Fischer.Whitehouse.ea:2010}.  The simulations are
evolved from core collapse, through the explosion up to 9~seconds
after bounce. Figure~\ref{fig:wind11} shows the evolution of the
luminosities and average energies for all neutrino flavors (left
panels).  The right panels show the evolution of the values of $Y_e$
and entropy asymptotically reached by the ejecta. One sees that the
early ejecta is neutron rich with $Y_e \sim 0.48$. This value is
larger than the one previously found in
ref.~\cite{Martinez-Pinedo.Fischer.ea:2012} using the TM1
EoS~\cite{Shen.Toki.ea:1998a}. The $Y_e$ values have been determined
using a full nuclear network that includes neutrino interactions both
on nucleons and nuclei and accounts for the so-called
$\alpha$-effect~\cite{Meyer.Mclaughlin.Fuller:1998}.

\begin{figure}[htb]
  \centering
  \includegraphics[width=\linewidth]{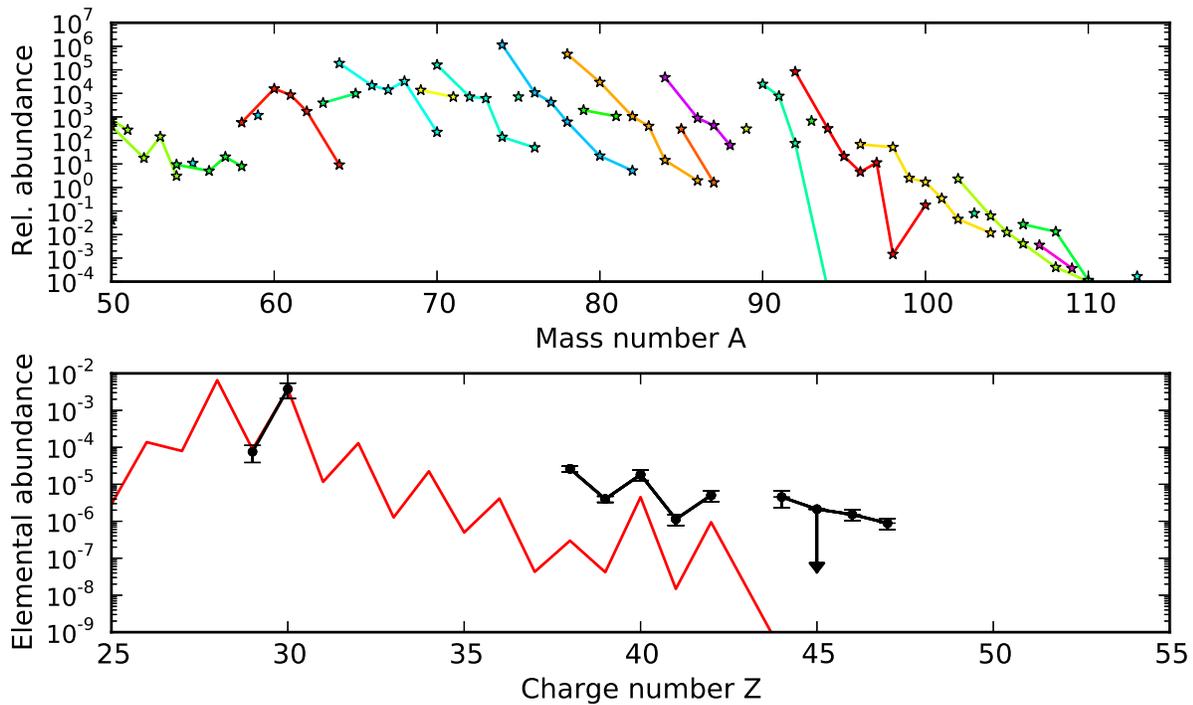}
  \caption{Mass-intergrated neutrino wind abundances. The upper panel
    shows the ratio between the abundance of a nucleus normalized to
    the solar abundance. The lower panel shows the elemental
    abundances (red) compared with the observations of
    HD~122563~\cite{Honda.Aoki.ea:2006}
    (black).  \label{fig:nucleosyn}} 
\end{figure}

The mass-integrated nucleosynthesis is shown in
figure~\ref{fig:nucleosyn}. The upper panel shows the mass-integrated
isotopic abundances normalized to the solar abundances. The lower
panel shows the mass-integrated elemental abundances compared with the
observations of the metal-poor star enriched in light r-process
elements HD~122563~\cite{Honda.Aoki.ea:2006}. The stellar observations
have been arbitrarily normalized to Zn ($Z=30$). Our calculations
reproduce the observed abundance of Zr ($Z=40$) and other nuclei
around $A=90$ within a factor 4. The production of these $N=50$ closed
neutron shell nuclei is rather sensitive to $Y_e$. They will be
overproduced if $Y_e\lesssim 0.47$~\cite{Hoffman.Woosley.ea:1996}. Our
results indicate that neutrino-driven winds are the site for the production
of elements like Sr, Y and Zr. This is in agreement with the
correlation observed in figure~\ref{fig:abundevol} as core-collapse
supernova are the main contributors for Fe at low
metallicities~\cite{Timmes.Woosley.Weaver:1995}.  In our calculations,
the elements Sr, Y, Zr, Nb, and Mo are produced mainly in the early
neutron-rich ejecta by charged-particle reactions together with some
neutron captures. Due to the sudden drop of alpha and neutron
separation energies around $N=50$ the production of nuclei with $N>50$
decreases dramatically (see upper panel
figure~\ref{fig:nucleosyn}). Nuclei with $Z>42$ ($A>92$) are mainly
produced in the late proton-rich ejecta by the $\nu
p$-process~\cite{Froehlich.Martinez-Pinedo.ea:2006,Pruet.Hoffman.ea:2006,Wanajo:2006}. However
their production is very inefficient due to the low antineutrino
luminosities at late times. The production of elements with $A>64$ by
the $\nu p$-process is very sensitive to the rate of antineutrino
absorption on protons~\cite{Martinez-Pinedo.Ziebarth.ea:2011} that is
directly proportional to the luminosity. Our value is around an order
of magnitude smaller than those used in
refs.~\cite{Arcones.Montes:2011,Wanajo:2013} but consistent with
recent Boltzmann transport
simulations~\cite{Huedepohl.Mueller.ea:2010,Roberts:2012}. In
addition, the amount of material ejected at later times is very small
to contribute significantly to the mass-integrated abundances.

If our results represent an standard situation for neutrino-driven  winds, the
obvious question then is: where are the elements with $Z>42$
(including Ag and Pd) produced? A recent
study~\cite{Hansen.Primas.ea:2012} has shown that the evolution with
metallicity of Ag and Pd is uncorrelated with both $N=50$ elements
(Sr, Y and Zr) and heavy r-process elements (Eu and Ba). Despite of
being consistent with our results, it represents a challenge for
nucleosynthesis models as it is very difficult to find astrophysical
conditions where Ag and Pd are produced without synthesizing neither
$N=50$ elements nor heavy r-process elements. This also applies to
electron-capture supernova that have been suggested as a possible site
for the production of elements between Ge and
Zr~\cite{Wanajo.Janka.Mueller:2011}. Another open problem that cannot
be explained by our simulations is the correlation between Eu and Fe
observed in figure~\ref{fig:abundevol}.

Due to the relatively large values of $Y_e$ achieved in our
nucleosynthesis calculations, only the neutron-deficient isotopes of
elements between Ge and Mo are produced. This  can be tested 
by observations of isotopic abundances in metal-poor stars. At the
same time it requires that the neutron-rich isotopes, that normally
cannot be produced by the s-process~\cite{Sneden.Cowan.Gallino:2008},
are produced in some other astrophysical site. Due to the more
neutron-rich conditions achieved in electron-capture
supernova~\cite{Wanajo.Janka.Mueller:2011} it is expected that their
distribution will be richer in neutron-rich isotopes. An alternative
site (but probably more speculative) is core-collapse supernova
explosions due to a quark-hadron phase
transition~\cite{Nishimura.Fischer.ea:2012}. It is interesting to
notice that $^{92}$Mo, a nucleus that is difficult to produce by the
p-process~\cite{Arnould.Goriely:2003}, is abundantly produced in our
calculations.

\section{Summary}
\label{sec:summary}

We have analyzed recent evidence of metal-poor stars that show the
existence of two distinctive sources for the production of r-process
nuclei with different frequencies along Galactic
history~\cite{Qian.Wasserburg:2007}. There is strong evidence that
core-collapse supernova could be the site for the production of light
r-process elements including Sr, Y and Zr (sometimes also denoted as
LEPP elements)~\cite{Arcones.Montes:2011}. They will account for the
abundances of elements observed in metal-poor stars that are enriched
in light r-process elements but depleted in heavy r-process
elements. HD~122563~\cite{Honda.Aoki.ea:2006,Roederer.Lawler.ea:2012}
constitutes the classical example. Looking at the evolution with
metallicity of several light r-process elements and Eu, that is taken
as a representative of heavy r-process elements, in HD~122563 type
stars we find a strong correlation. It indicates that in fact the
abundances of Sr, Y and Zr seen in these stars are due to
core-collapse supernova that is the major producer of iron at low
metallicity. The situation for Ag is unclear as only two of the stars
that survive our selection criteria have observed abundances of Ag. A
recent study has in fact shown that the abundance of Ag is
anticorrelated with Sr, Y and
Zr~\cite{Hansen.Primas.ea:2012}. Surprisingly, we also find that the
Eu abundance observed on HD~122563 type stars is correlated with
Fe. It will be interesting to see if this correlation extends to other
elements like Ba.

We have presented first long-term Boltzmann transport neutrino-wind
simulations based on a core-collapse supernova equation of state that
reproduces recent constrains on the nuclear symmetry energy at
saturation densities~\cite{Lattimer.Lim:2013,Krueger.Tews.ea:2013} and
on the energy per nucleon of neutron
matter~\cite{Krueger.Tews.ea:2013}. Our simulations consistently
include mean-field corrections for neutrons and
protons~\cite{Reddy.Prakash.Lattimer:1998,Martinez-Pinedo.Fischer.ea:2012,Roberts.Reddy.Shen:2012}
for the (anti)neutrino opacities at high density. These corrections
increase sustantially the opacity for $\nu_e$. The increase being
largest the larger the mean-field energy difference between neutrons
and protons, i.e. the larger the symmetry energy of the EoS. A larger
symmetry energy implies a larger difference between the average
energies of the emitted $\bar{\nu}_e$ and
$\nu_e$~\cite{Martinez-Pinedo.Fischer.ea:2012,Roberts.Reddy.Shen:2012,Horowitz.Shen.ea:2012}
and a smaller neutrino
luminosity~\cite{Roberts.Shen.ea:2012,Martinez-Pinedo.Fischer.ea:2012}. Both
aspects are important for determining the nucleosynthesis in
neutrino-driven winds.

Our simulations predict that the early neutrino-driven wind ejecta is
neutron-rich with $Y_e \approx 0.48$. At times around 4~s the ejecta
become proton-rich reaching $Y_e = 0.586$ at 9~s when our simulation
stops. We have analyzed the nucleosynthesis outcome of the
simulations. Our results show that neutrino-driven winds do not
produce heavy r-process elements. They contribute to the production of
elements between Zn and Mo, including $^{92}$Mo that is largely
underproduced by the p-process in massive
stars~\cite{Arnould.Goriely:2003}. The elemental abundances are
consistent with the observations of the metal-poor star HD~12263. For
elements between Ge and Mo, we produce mainly the neutron-deficient
isotopes. This prediction could be confirmed by observations of
isotopic abundances in metal-poor stars. No elements heavier than Mo
($Z=42$) are produced in our calculations. In neutron-rich ejecta,
their production is suppressed due to the $N=50$ shell closure. In the
late proton-rich ejecta, the $\nu p$-process, that in principle can
contribute to their production, is very inefficient due to the low
antineutrino luminosities. 

Even if our nucleosynthesis study is based on parameter free Boltzmann
transport simulations, there are still some sources of uncertainty on
the determination of the (anti)neutrino opacities. One is the
treatment of neutrino interactions with light nuclear clusters, in
particular $^2$H, $^3$H. They are expected to be present in the region
of neutrino decoupling~\cite{Arcones.Martinez-Pinedo.ea:2008} and in
fact the equation of state we use predicts them with large abundances
(see figure 10 of ref.~\cite{Fischer.Hempel.ea:2013}). The presence of
light nuclear clusters affects the (anti)neutrino opacities in
different ways. First, it modifies the mean-field corrections for the
unbound nucleons~\cite{Horowitz.Shen.ea:2012}. This effect is included
in our simulations. Secondly, it becomes necessary to consider both
charged and neutral current interactions with the light clusters. The
relevant cross sections have been computed for the
deuteron~\cite{Nakamura.Sato.ea:2001,Nakamura.Sato.ea:2002,Shen.Marcucci.ea:2012}
and triton~\cite{Arcones.Martinez-Pinedo.ea:2008} in vacuum. However,
their presence in the nuclear medium will induce changes in the energy
thresholds for the reactions due to different in medium modifications
for the for the initial and final states. Furthermore, the wave
functions~\cite{Roepke.Bastian.ea:2013} for the cluster will
be modified with respect to the vacuum case affecting the response to
(anti)neutrinos.

An accurate determination of the (anti)neutrino spectra and
luminosities and their evolution with time is also necessary for
neutrino flavor oscillations in
supernova~\cite{Duan.Fuller.Qian:2010,Duan.Kneller:2009} and neutrino
detection on Earth. From the point of view of nucleosynthesis in
neutrino-driven winds collective neutrino
oscillations~\cite{Duan.Fuller.Qian:2010} are most relevant. It
remains to evaluate the impact that they may have on the
nucleosynthesis. Another important effect is the possibility of
oscillations of active to sterile neutrinos, suggested to explain the
reactor anomaly~\cite{Kopp.Machado.ea:2013}. Oscillations to the
sterile flavor can affect both the dynamics during the supernova
explosion~\cite{Nunokawa.Peltoniemi.ea:1997,Wu.Fischer.ea:2013} and
the neutrino-driven wind
nucleosynthesis~\cite{Tamborra.Raffelt.ea:2012,Wu.Fischer.ea:2013}.

\ack

G.M.P. and L.H. is partly supported by the Deutsche
Forschungsgemeinschaft through contract SFB 634, the Helmholtz
International Center for FAIR within the framework of the LOEWE
program launched by the state of Hesse, and the Helmholtz Association
through the Nuclear Astrophysics Virtual Institute (VH-VI-417). T.F.
acknowledges support from the Narodowe Centrum Nauki (NCN) within the
``Maestro'' program under contract No. DEC- 2011/02/A/ST2/00306. We
gratefully acknowledge Stefan Typel and Thomas Kr\"uger for providing
the data for figure~\ref{fig:eos}. The supernova simulations were
performed at the computer cluster at the GSI Helmholtzzentrum for
Schwerionenforschung GmbH, Darmstadt (Germany). 

\section*{References}


\begin{thebibliography}{100}
\providecommand{\url}[1]{\texttt{#1}}
\providecommand{\eprint}[2][]{\texttt{arXiv:#2~[#1]}}

\bibitem{Bethe:1990}
Bethe H~A 1990 \emph{Rev. Mod. Phys.} \textbf{62} 801

\bibitem{Janka.Langanke.ea:2007}
Janka H~T, Langanke K, Marek A, Mart{\'i}nez-Pinedo G and M{\"u}ller B 2007
  \emph{Phys. Repts.} \textbf{442} 38

\bibitem{Bethe.Wilson:1985}
Bethe H~A and Wilson J~R 1985 \emph{Astrophys. J.} \textbf{295} 14

\bibitem{Mueller.Janka.Marek:2012}
{M{\"u}ller} B, {Janka} H~T and {Marek} A 2012 \emph{Astrophys. J.}
  \textbf{756} 84

\bibitem{Bruenn.Mezzacappa.ea:2013}
{Bruenn} S~W, {Mezzacappa} A, {Hix} W~R, {Lentz} E~J, {Bronson Messer} O~E,
  {Lingerfelt} E~J, {Blondin} J~M, {Endeve} E, {Marronetti} P and {Yakunin} K~N
  2013 \emph{Astrophys. J.} \textbf{767} L6

\bibitem{Suwa.Takiwaki.ea:2013}
{Suwa} Y, {Takiwaki} T, {Kotake} K, {Fischer} T, {Liebend{\"o}rfer} M and
  {Sato} K 2013 \emph{Astrophys. J.} \textbf{764} 99

\bibitem{duncan.shapiro.wasserman:1986}
{Duncan} R~C, {Shapiro} S~L and {Wasserman} I 1986 \emph{Astrophys. J.}
  \textbf{309} 141

\bibitem{Arcones.Thielemann:2013}
Arcones A and Thielemann F~K 2013 \emph{J. Phys. G: Nucl. Part. Phys.}
  \textbf{40} 013201

\bibitem{Duan.Kneller:2009}
Duan H and Kneller J~P 2009 \emph{J. Phys. G: Nucl. Part. Phys.} \textbf{36}
  113201

\bibitem{Duan.Fuller.Qian:2010}
{Duan} H, {Fuller} G~M and {Qian} Y 2010 \emph{Ann. Rev. Nucl. Part. Sci.}
  \textbf{60} 569

\bibitem{Woosley.Hartmann.ea:1990}
{Woosley} S~E, {Hartmann} D~H, {Hoffman} R~D and {Haxton} W~C 1990
  \emph{Astrophys. J.} \textbf{356} 272

\bibitem{Heger.Kolbe.ea:2005}
Heger A, Kolbe E, Haxton W, Langanke K, Mart{\'\i}nez-Pinedo G and Woosley S~E
  2005 \emph{Phys. Lett. B} \textbf{606} 258

\bibitem{Banerjee.Haxton.Qian:2011}
Banerjee P, Haxton W~C and Qian Y~Z 2011 \emph{Phys. Rev. Lett.} \textbf{106}
  201104

\bibitem{Burbidge.Burbidge.ea:1957}
{Burbidge} E~M, {Burbidge} G~R, {Fowler} W~A and {Hoyle} F 1957 \emph{Rev. Mod.
  Phys.} \textbf{29} 547

\bibitem{Cameron:1957}
Cameron A~G~W 1957 Stellar evolution, nuclear astrophysics, and nucleogenesis
  Report CRL-41 Chalk River

\bibitem{Arnould.Goriely:2003}
{Arnould} M and {Goriely} S 2003 \emph{Phys. Repts.} \textbf{384} 1

\bibitem{straniero.gallino.cristallo:2006}
Straniero O, Gallino R and Cristallo S 2006 \emph{Nucl. Phys. A} \textbf{777}
  311

\bibitem{Busso.Gallino.Wasserburg:1999}
Busso M, Gallino R and Wasserburg G~J 1999 \emph{Annu. Rev. Astron. Astrophys.}
  \textbf{37} 239

\bibitem{Pignatari.Gallino.ea:2010}
{Pignatari} M, {Gallino} R, {Heil} M, {Wiescher} M, {K{\"a}ppeler} F, {Herwig}
  F and {Bisterzo} S 2010 \emph{Astrophys. J.} \textbf{710} 1557

\bibitem{arnould.goriely.takahashi:2007}
Arnould M, Goriely S and Takahashi K 2007 \emph{Phys. Repts.} \textbf{450} 97

\bibitem{Lodders:2003}
{Lodders} K 2003 \emph{Astrophys. J.} \textbf{591} 1220

\bibitem{Asplund.Grevesse.ea:2009}
{Asplund} M, {Grevesse} N, {Sauval} A~J and {Scott} P 2009 \emph{Ann. Rev.
  Astron. Astrophys.} \textbf{47} 481

\bibitem{Jacobson.Frebel:2013}
{Jacobson} H~R and {Frebel} A 2013  \eprint[astro-ph.GA]{1309.0037}

\bibitem{Arlandini.Kaeppeler.ea:1999}
{Arlandini} C, {K{\"a}ppeler} F, {Wisshak} K, {Gallino} R, {Lugaro} M, {Busso}
  M and {Straniero} O 1999 \emph{Astrophys. J.} \textbf{525} 886

\bibitem{Bisterzo.Gallino.ea:2010}
{Bisterzo} S, {Gallino} R, {Straniero} O, {Cristallo} S and {K{\"a}ppeler} F
  2010 \emph{Mon. Not. Roy. Ast. Soc.} \textbf{404} 1529

\bibitem{Cowan.Sneden:2006}
Cowan J~J and Sneden C 2006 \emph{Nature} \textbf{440} 1151

\bibitem{Sneden.Cowan.Gallino:2008}
{Sneden} C, {Cowan} J~J and {Gallino} R 2008 \emph{Ann. Rev. Astron.
  Astrophys.} \textbf{46} 241

\bibitem{Christlieb.Beers.ea:2004}
{Christlieb} N \emph{et~al.} 2004 \emph{Astron. \& Astrophys.} \textbf{428}
  1027

\bibitem{Roederer.Lawler:2012}
{Roederer} I~U and {Lawler} J~E 2012 \emph{Astrophys. J.} \textbf{750} 76

\bibitem{Honda.Aoki.ea:2006}
{Honda} S, {Aoki} W, {Ishimaru} Y, {Wanajo} S and {Ryan} S~G 2006
  \emph{Astrophys. J.} \textbf{643} 1180

\bibitem{Roederer.Lawler.ea:2012}
{Roederer} I~U, {Lawler} J~E, {Sobeck} J~S, {Beers} T~C, {Cowan} J~J, {Frebel}
  A, {Ivans} I~I, {Schatz} H, {Sneden} C and {Thompson} I~B 2012
  \emph{Astrophys. J. Suppl.} \textbf{203} 27

\bibitem{Woosley.Hoffman:1992}
Woosley S~E and Hoffman R~D 1992 \emph{Astrophys. J.} \textbf{395} 202

\bibitem{Witti.Janka.Takahashi:1994}
Witti J, Janka H~T and Takahashi K 1994 \emph{Astron. \& Astrophys.}
  \textbf{286} 841

\bibitem{Qian.Wasserburg:2007}
{Qian} Y~Z and {Wasserburg} G~J 2007 \emph{Phys. Repts.} \textbf{442} 237

\bibitem{Travaglio.Gallino.ea:2004}
{Travaglio} C, {Gallino} R, {Arnone} E, {Cowan} J, {Jordan} F and {Sneden} C
  2004 \emph{Astrophys. J.} \textbf{601} 864

\bibitem{Burris.Pilachowski.ea:2000}
{Burris} D~L, {Pilachowski} C~A, {Armandroff} T~E, {Sneden} C, {Cowan} J~J and
  {Roe} H 2000 \emph{Astrophys. J.} \textbf{544} 302

\bibitem{Honda.Aoki.ea:2004}
{Honda} S, {Aoki} W, {Kajino} T, {Ando} H, {Beers} T~C, {Izumiura} H,
  {Sadakane} K and {Takada-Hidai} M 2004 \emph{Astrophys. J.} \textbf{607} 474

\bibitem{Barklem.Christlieb.ea:2005}
{Barklem} P~S, {Christlieb} N, {Beers} T~C, {Hill} V, {Bessell} M~S, {Holmberg}
  J, {Marsteller} B, {Rossi} S, {Zickgraf} F~J and {Reimers} D 2005
  \emph{Astron. \& Astrophys.} \textbf{439} 129

\bibitem{Hansen.Primas.ea:2012}
{Hansen} C~J, {Primas} F, {Hartman} H, {Kratz} K~L, {Wanajo} S, {Leibundgut} B,
  {Farouqi} K, {Hallmann} O, {Christlieb} N and {Nilsson} H 2012 \emph{Astron.
  \& Astrophys.} \textbf{545} A31

\bibitem{Cowan.Thielemann:2004}
Cowan J~J and Thielemann F~K 2004 \emph{Physics Today} \textbf{57} 47

\bibitem{Wasserburg.Qian:2000}
Wasserburg G~J and Qian Y~Z 2000 \emph{Astrophys. J.} \textbf{529} L21

\bibitem{Roederer:2013}
{Roederer} I~U 2013 \emph{Astron. J.} \textbf{145} 26

\bibitem{Montes.Beers.ea:2007}
Montes F \emph{et~al.} 2007 \emph{Astrophys. J.} \textbf{671} 1685

\bibitem{Korobkin.Rosswog.ea:2012}
{Korobkin} O, {Rosswog} S, {Arcones} A and {Winteler} C 2012 \emph{Mon. Not.
  Roy. Ast. Soc.} \textbf{426} 1940

\bibitem{Bauswein.Goriely.Janka:2013}
Bauswein A, Goriely S and Janka H~T 2013 \emph{Astrophys. J.} \textbf{773} 78

\bibitem{Martinez-Pinedo:2008}
{Mart{\'{\i}}nez-Pinedo} G 2008 \emph{J. Phys. G: Nucl. Part. Phys.}
  \textbf{35} 014057

\bibitem{Petermann.Langanke.ea:2012}
{Petermann} I, {Langanke} K, {Mart{\'{\i}}nez-Pinedo} G, {Panov} I~V,
  {Reinhard} P~G and {Thielemann} F~K 2012 \emph{Eur. Phys. J. A} \textbf{48}
  122

\bibitem{Metzger.Martinez-Pinedo.ea:2010}
{Metzger} B~D, {Mart{\'{\i}}nez-Pinedo} G, {Darbha} S, {Quataert} E, {Arcones}
  A, {Kasen} D, {Thomas} R, {Nugent} P, {Panov} I~V and {Zinner} N~T 2010
  \emph{Mon. Not. Roy. Ast. Soc.} \textbf{406} 2650

\bibitem{Roberts.Kasen.ea:2011}
{Roberts} L~F, {Kasen} D, {Lee} W~H and {Ramirez-Ruiz} E 2011 \emph{Astrophys.
  J.} \textbf{736} L21

\bibitem{Tanvir.Levan.ea:2013}
{Tanvir} N~R, {Levan} A~J, {Fruchter} A~S, {Hjorth} J, {Hounsell} R~A,
  {Wiersema} K and {Tunnicliffe} R~L 2013 \emph{Nature} \textbf{500} 547

\bibitem{Berger.Fong.Chornock:2013}
{Berger} E, {Fong} W and {Chornock} R 2013 \emph{Astrophys. J.} \textbf{774}
  L23

\bibitem{Grossman.Korobkin.ea:2013}
{Grossman} D, {Korobkin} O, {Rosswog} S and {Piran} T 2013
  \eprint[astro-ph.HE]{1307.2943}

\bibitem{Wanajo.Janka:2012}
{Wanajo} S and {Janka} H~T 2012 \emph{Astrophys. J.} \textbf{746} 180

\bibitem{Fernandez.Metzger:2013}
{Fern{\'a}ndez} R and {Metzger} B~D 2013 \emph{Mon. Not. Roy. Ast. Soc.}
  \eprint[astro-ph.HE]{1304.6720}
  \url{http://dx.doi.org/10.1093/mnras/stt1312}

\bibitem{Qian:2000}
Qian Y~Z 2000 \emph{Astrophys. J.} \textbf{534} L67

\bibitem{Argast.Samland.ea:2004}
{Argast} D, {Samland} M, {Thielemann} F~K and {Qian} Y~Z 2004 \emph{Astron. \&
  Astrophys.} \textbf{416} 997

\bibitem{Winteler.Kaeppeli.ea:2012}
{Winteler} C, {K{\"a}ppeli} R, {Perego} A, {Arcones} A, {Vasset} N, {Nishimura}
  N, {Liebend{\"o}rfer} M and {Thielemann} F~K 2012 \emph{Astrophys. J.}
  \textbf{750} L22

\bibitem{Woosley.Wilson.ea:1994}
Woosley S~E, Wilson J~R, Mathews G~J, Hoffman R~D and Meyer B~S 1994
  \emph{Astrophys. J.} \textbf{433} 229

\bibitem{Takahashi.Witti.Janka:1994}
Takahashi K, Witti J and Janka H~T 1994 \emph{Astron. \& Astrophys.}
  \textbf{286} 857

\bibitem{Qian.Woosley:1996}
{Qian} Y~Z and {Woosley} S~E 1996 \emph{Astrophys. J.} \textbf{471} 331

\bibitem{hoffman.woosley.qian:1997}
{Hoffman} R~D, {Woosley} S~E and {Qian} Y~Z 1997 \emph{Astrophys. J.}
  \textbf{482} 951

\bibitem{Otsuki.Tagoshi.ea:2000}
Otsuki K, Tagoshi H, Kajino T and Wanajo S 2000 \emph{Astrophys. J.}
  \textbf{533} 424

\bibitem{Thompson.Burrows.Meyer:2001}
Thompson T~A, Burrows A and Meyer B~S 2001 \emph{Astrophys. J.} \textbf{562}
  887

\bibitem{Arcones.Janka.Scheck:2007}
{Arcones} A, {Janka} H~T and {Scheck} L 2007 \emph{Astron. \& Astrophys.}
  \textbf{467} 1227

\bibitem{Arcones.Janka:2011}
{Arcones} A and {Janka} H~T 2011 \emph{Astron. \& Astrophys.} \textbf{526} A160

\bibitem{Kuroda.Wanajo.Nomoto:2008}
{Kuroda} T, {Wanajo} S and {Nomoto} K 2008 \emph{Astrophys. J.} \textbf{672}
  1068

\bibitem{Roberts.Woosley.Hoffman:2010}
{Roberts} L~F, {Woosley} S~E and {Hoffman} R~D 2010 \emph{Astrophys. J.}
  \textbf{722} 954

\bibitem{Arcones.Montes:2011}
{Arcones} A and {Montes} F 2011 \emph{Astrophys. J.} \textbf{731} 5

\bibitem{Martinez-Pinedo.Fischer.ea:2012}
Mart{\'\i}nez-Pinedo G, Fischer T, Lohs A and Huther L 2012 \emph{Phys. Rev.
  Lett.} \textbf{109} 251104

\bibitem{Roberts:2012}
{Roberts} L~F 2012 \emph{Astrophys. J.} \textbf{755} 126

\bibitem{Roberts.Reddy.Shen:2012}
{Roberts} L~F, {Reddy} S and {Shen} G 2012 \emph{Phys. Rev. C} \textbf{86}
  065803

\bibitem{Wanajo:2013}
Wanajo S 2013 \emph{Astrophys. J.} \textbf{770} L22

\bibitem{keil.raffelt.janka:2003}
{Keil} M~T, {Raffelt} G~G and {Janka} H~T 2003 \emph{Astrophys. J.}
  \textbf{590} 971

\bibitem{Froehlich.Hauser.ea:2006}
Fr{\"o}hlich C \emph{et~al.} 2006 \emph{Astrophys. J.} \textbf{637} 415

\bibitem{Martinez-Pinedo:2008b}
{Mart{\'{\i}}nez-Pinedo} G 2008 \emph{Eur. Phys. J. Special Topics}
  \textbf{156} 123

\bibitem{Buras.Rampp.ea:2006}
Buras R, Rampp M, Janka H~T and Kifonidis K 2006 \emph{Astron. \& Astrophys.}
  \textbf{447} 1049

\bibitem{Liebendoerfer.Mezzacappa.ea:2001a}
Liebend{\"o}rfer M, Mezzacappa A, Thielemann F~K, {Bronson Messer} O~E,
  {Raphael Hix} W and Bruenn S~W 2001 \emph{Phys. Rev. D} \textbf{63} 103004

\bibitem{Thompson.Quataert.Burrows:2005}
{Thompson} T~A, {Quataert} E and {Burrows} A 2005 \emph{Astrophys. J.}
  \textbf{620} 861

\bibitem{Huedepohl.Mueller.ea:2010}
{H{\"u}depohl} L, {M{\"u}ller} B, {Janka} H, {Marek} A and {Raffelt} G~G 2010
  \emph{Phys. Rev. Lett.} \textbf{104} 251101

\bibitem{Fischer.Whitehouse.ea:2010}
Fischer T, Whitehouse S~C, Mezzacappa A, Thielemann F~K and Liebend\"orfer M
  2010 \emph{Astron. \& Astrophys.} \textbf{517} A80

\bibitem{Fischer.Martinez-Pinedo.ea:2012}
Fischer T, Mart{\'i}nez-Pinedo G, Hempel M and Liebend{\"o}rfer M 2012
  \emph{Phys. Rev. D} \textbf{85} 083003

\bibitem{Raffelt:2001}
Raffelt G~G 2001 \emph{Astrophys. J.} \textbf{561} 890

\bibitem{Reddy.Prakash.Lattimer:1998}
Reddy S, Prakash M and Lattimer J~M 1998 \emph{Phys. Rev. D} \textbf{58} 013009

\bibitem{Lattimer.Swesty:1991}
Lattimer J~M and Swesty F~D 1991 \emph{Nucl. Phys. A} \textbf{535} 331

\bibitem{Shen.Toki.ea:1998a}
{Shen} H, {Toki} H, {Oyamatsu} K and {Sumiyoshi} K 1998 \emph{Nucl. Phys. A}
  \textbf{637} 435

\bibitem{Haensel.Potekhin.Yakovlev:2007}
Haensel P, Potekhin A~Y and Yakovlev D~G 2007 \emph{Neutron Stars 1: Equation
  of State and Structure} volume 326 of \emph{Astrophysics and Space Science
  Library} (Springer, New York)

\bibitem{Bruenn:1985}
Bruenn S~W 1985 \emph{Astrophys. J. Suppl.} \textbf{58} 771

\bibitem{Roberts.Shen.ea:2012}
Roberts L~F, Shen G, Cirigliano V, Pons J~A, Reddy S and Woosley S~E 2012
  \emph{Phys. Rev. Lett.} \textbf{108} 061103

\bibitem{Langanke.Martinez-Pinedo:2003}
Langanke K and Mart{\'\i}nez-Pinedo G 2003 \emph{Rev. Mod. Phys.} \textbf{75}
  819

\bibitem{martinez-pinedo.liebendoerfer.frekers:2006}
Mart{\'\i}nez-Pinedo G, Liebend{\"o}rfer M and Frekers D 2006 \emph{Nucl. Phys.
  A} \textbf{777} 395

\bibitem{Mueller.Janka.Heger:2012}
{M{\"u}ller} B, {Janka} H~T and {Heger} A 2012 \emph{Astrophys. J.}
  \textbf{761} 72

\bibitem{Tews.Krueger.ea:2013}
Tews I, Kr\"uger T, Hebeler K and Schwenk A 2013 \emph{Phys. Rev. Lett.}
  \textbf{110} 032504

\bibitem{Krueger.Tews.ea:2013}
Kr\"uger T, Tews I, Hebeler K and Schwenk A 2013 \emph{Phys. Rev. C}
  \textbf{88} 025802

\bibitem{Lattimer.Lim:2013}
{Lattimer} J~M and {Lim} Y 2013 \emph{Astrophys. J.} \textbf{771} 51

\bibitem{Typel.Roepke.ea:2010}
Typel S, R\"opke G, Kl\"ahn T, Blaschke D and Wolter H~H 2010 \emph{Phys. Rev.
  C} \textbf{81} 015803

\bibitem{horowitz.schwenk:2006}
Horowitz C~J and Schwenk A 2006 \emph{Nucl. Phys. A} \textbf{776} 55

\bibitem{Fischer.Hempel.ea:2013}
{Fischer} T, {Hempel} M, {Sagert} I, {Suwa} Y and {Schaffner-Bielich} J 2013
  \eprint[astro-ph.HE]{1307.6190}

\bibitem{Hempel.Schaffner-Bielich:2010}
Hempel M and Schaffner-Bielich J 2010 \emph{Nucl. Phys. A} \textbf{837} 210

\bibitem{Woosley.Heger.Weaver:2002}
Woosley S~E, Heger A and Weaver T~A 2002 \emph{Rev. Mod. Phys.} \textbf{74}
  1015

\bibitem{Meyer.Mclaughlin.Fuller:1998}
{Meyer} B~S, {McLaughlin} G~C and {Fuller} G~M 1998 \emph{Phys. Rev. C}
  \textbf{58} 3696

\bibitem{Hoffman.Woosley.ea:1996}
{Hoffman} R~D, {Woosley} S~E, {Fuller} G~M and {Meyer} B~S 1996
  \emph{Astrophys. J.} \textbf{460} 478

\bibitem{Timmes.Woosley.Weaver:1995}
Timmes F~X, Woosley S~E and Weaver T~A 1995 \emph{Astrophys. J. Suppl.}
  \textbf{98} 617

\bibitem{Froehlich.Martinez-Pinedo.ea:2006}
Fr{\"o}hlich C, Mart{\'\i}nez-Pinedo G, Liebend{\"o}rfer M, Thielemann F~K,
  Bravo E, Hix W~R, Langanke K and Zinner N~T 2006 \emph{Phys. Rev. Lett.}
  \textbf{96} 142502

\bibitem{Pruet.Hoffman.ea:2006}
Pruet J, Hoffman R~D, Woosley S~E, Janka H~T and Buras R 2006 \emph{Astrophys.
  J.} \textbf{644} 1028

\bibitem{Wanajo:2006}
Wanajo S 2006 \emph{Astrophys. J.} \textbf{647} 1323

\bibitem{Martinez-Pinedo.Ziebarth.ea:2011}
Mart{\'i}nez-Pinedo G, Ziebarth B, Fischer T and Langanke K 2011 \emph{Eur.
  Phys. J. A} \textbf{47} 1

\bibitem{Wanajo.Janka.Mueller:2011}
Wanajo S, Janka H~T and M{\"u}ller B 2011 \emph{Astrophys. J.} \textbf{726} L15

\bibitem{Nishimura.Fischer.ea:2012}
Nishimura N, Fischer T, Thielemann F~K, Fr{\"o}lich C, Hempel M, K{\"a}ppeli R,
  Mart{\'i}nez-Pinedo G, Rauscher T, Sagert I and Winteler C 2012
  \emph{Astrophys. J.} \textbf{758} 9

\bibitem{Horowitz.Shen.ea:2012}
Horowitz C~J, Shen G, O'Connor E and Ott C~D 2012 \emph{Phys. Rev. C}
  \textbf{86} 065806

\bibitem{Arcones.Martinez-Pinedo.ea:2008}
Arcones A, {Mart{\'i}nez-Pinedo} G, {O'Connor} E, Schwenk A, Janka H, Horowitz
  C~J and Langanke K 2008 \emph{Phys. Rev. C} \textbf{78} 015806

\bibitem{Nakamura.Sato.ea:2001}
Nakamura S, Sato T, Gudkov V and Kubodera K 2001 \emph{Phys. Rev. C}
  \textbf{63} 034617

\bibitem{Nakamura.Sato.ea:2002}
Nakamura S, Sato T, Ando S, Park T~S, Myhrer F, Gudkov V and Kubodera K 2002
  \emph{Nucl. Phys. A} \textbf{707} 561

\bibitem{Shen.Marcucci.ea:2012}
Shen G, Marcucci L~E, Carlson J, Gandolfi S and Schiavilla R 2012 \emph{Phys.
  Rev. C} \textbf{86} 035503

\bibitem{Roepke.Bastian.ea:2013}
R{\o}pke G, Bastian N~U, Blaschke D, Kl{\"a}hn T, Typel S and Wolter H 2013
  \emph{Nucl. Phys. A} \textbf{897} 70

\bibitem{Kopp.Machado.ea:2013}
{Kopp} J, {Machado} P~A~N, {Maltoni} M and {Schwetz} T 2013 \emph{Journal of
  High Energy Physics} \textbf{5} 50 \eprint[hep-ph]{1303.3011}

\bibitem{Nunokawa.Peltoniemi.ea:1997}
Nunokawa H, Peltoniemi J~T, Rossi A and Valle J~W~F 1997 \emph{Phys. Rev. D}
  \textbf{56} 1704

\bibitem{Wu.Fischer.ea:2013}
{Wu} M~R, {Fischer} T, {Mart{\'{\i}}nez-Pinedo} G and {Qian} Y~Z 2013
  \eprint[astro-ph.HE]{1305.2382}

\bibitem{Tamborra.Raffelt.ea:2012}
Tamborra I, Raffelt G~G, H{\"u}depohl L and Janka H~T 2012 \emph{JCAP}
  \textbf{01} 013

\end{thebibliography}

\end{document}